# On the oxidation state of titanium in titanium dioxide

Daniel Koch, Sergei Manzhos[1]

Department of Mechanical Engineering, National University of Singapore, 9 Engineering Drive 1, Singapore 117576, Singapore

**Abstract**

The oxidation state of titanium in titanium dioxide is commonly assumed to be +4. This assumption is used ubiquitously to rationalize phenomena observed with $TiO_2$. We present a comprehensive electronic structure investigation of Ti ions, $TiO_2$ molecules and $TiO_2$ bulk crystals, using different density functional theory and wave function-based approaches, which suggests a lower oxidation state (+3). Specifically, there is evidence of a significant remaining contribution from valence *s* and *d* electrons of Ti, including the presence of a nuclear cusp around the Ti core. The charge corresponding to valence *s* and *d* states of Ti amounts to 1 *e*. The commonly assumed picture may therefore have to be revised.

**1. Introduction**

Titanium dioxide ($TiO_2$) is a material widely used in numerous technologies. $TiO_2$ polymorphs are large-band-gap semiconductors which makes them suitable for applications in photocatalysis and solar cells.[1,2] The excitation energy for the formation of an electron-hole pair in pure titanium dioxides is in the energy range of ultraviolet light, while doping also makes photoexcitation with visible light accessible.[3] Besides photocatalysis, the $TiO_2$ polymorphs are promising electrode materials for Li-ion and Na-ion batteries, due to their potentially high capacity, cycling stability and charge/discharge rate.[4,5] The band structure of titania compounds was reported extensively, commonly describing the valence band as being primarily composed of O 2*p* states and and the conduction band mostly consisting of Ti 3*d* contributions.[6,7] At the same time, the density of states shows other less important contributions, notably from Ti 3*d* states in the valence band.[8,9]

The commonly accepted picture of the ionic oxidation states in $TiO_2$ is a quadruply charged $Ti^{4+}$ cation held together with $O^{2-}$ anions by highly ionic bonds, an assumption based on qualitative MO considerations.[10,11] Indeed, the often-cited basis for the assumption of $Ti^{4+}$, photoelectron spectroscopy, relied on these very consideration to interpret XPS peaks.[12,13] There does not seem to be direct, independent experimental evidence of the absolute oxidation state. More recent quantum chemical computations, on the other hand, predict the mentioned $3d$ contributions to the $TiO_2$ valence band and an additional small fraction of $4s$ states in periodic titania systems. Furthermore, several charge analysis techniques such as Mulliken[14] or Bader[15] charge analysis predict a significant charge remainder at the Ti centers in $TiO_2$ polymorphs, suggesting Ti-O bonds with stronger covalent character.[16,17] The computed charge states significantly lower than +4 are usually assumed to result from the (imperfect) definitions of charge transfer measures and have not led to questioning of the assumed $Ti^{4+}$ oxidation state. This assumption is used ubiquitously to rationalize phenomena observed with $TiO_2$, such as doping or Li/Na insertion which are said to lead to the formation of $Ti^{3+}$ states.[18,19] The aim of this work is to analyze the electronic structure of titania, systematically compare different charge measures on different titania systems and to quantify the remaining charge on the Ti centers in those systems.

---

[1] Author to whom correspondance should be addressed. E-mail: mpemanzh@nus.edu.sg . Tel: +65 6516 4605; fax: +65 6779 1459.

From Kato's cusp theorem[20] follows that the spherically averaged electron density $\bar{\rho}(r)$ at a nucleus A in a many-electron system exhibits a cusp and at the nuclear position with $r_A=0$ the following relation holds:

$$\lim_{r_A \to 0} \left( \frac{\partial}{\partial r_A} + 2 Z_A \right) \bar{\rho}(r_A) = 0 \qquad (1)$$

with $Z_A$ being the charge of nucleus A. However, this cusp condition is only correct for wave functions composed of Slater-type orbitals and their resulting densities. Densities constructed from Gaussian-type basis functions do not exhibit a cusp, but rather a maximum of the all-electron ground state at the nuclear position. This maximum occurs due to significant $s$-contributions always present in many-electron wave functions.

For valence densities of mixed $s$- and higher angular momentum contributions, the occurrence of a maximum or minimum in the spherically averaged density at r=0 is dependent on the relation of the prefactors of different basis function types and the exponential coefficients of the $s$-type functions. However, the occurrence of a maximum at r=0 indicates in every case a non-negligible contribution of $s$-type functions centered at the atom in a single-nucleus system. Since higher order $s$-, $p$-, $d$- and $f$-functions exhibit radial maxima, the assignment of nuclear maxima is ambiguous in molecular systems using a Gaussian-type basis. In the case of Slater-type basis functions, the occurrence of a cusp indicates a density remainder at the atom stemming from atom-centered function at this nucleus, while this assumption cannot be made for Gaussian-type orbitals anymore and has then to be verified by other means.

In the framework of density functional theory (DFT), the spherically averaged density of a many-electron system is the sum of spherically averaged densities of all orbitals, weighted by occupancy. By taking into account only valence orbitals, a spherically averaged valence density can be in principle used to analyze the valence density distribution around an atom. Since the DFT solutions for orbitals and orbital eigenenergies are not necessarily related to the true one-electron properties of the system, these have to be used with caution.

Several methods for the quantification of charges on distinct centers in polyatomic compounds exist in the framework of electronic structure theory. One of the first proposed was the Mulliken population analysis[14] which uses entrywise products of density and overlap matrix $\mathbf{D} \circ \mathbf{S}$, expressing the remaining charge density on a center A as

$$\rho_A = \sum_{\alpha \in A} \sum_{\beta} D_{\alpha\beta} S_{\alpha\beta} \qquad (2)$$

and the corresponding atomic charge $Q_A$ as

$$Q_A = Z_A - \rho_A \quad . \qquad (3)$$

Since the population distribution between two centers is arbitrarily set as equal and the results are basis set dependent, further approaches were proposed over time, like the Hirshfeld analysis.[21-23] In it, the ratio of spherically averaged density of the atom of concern A and the sum of all atomic densities is used as weighting factor $w_A(r)$

$$w_A(r) = \frac{\overline{\rho_A}(r)}{\sum_K \overline{\rho_K}(r)} \qquad (4)$$

and the atomic charge is calculated from the molecular density $\rho_{mol}(r)$ as

$$Q_A = Z_A - \int_{r=0}^{\infty} w_A(r) \rho_{mol}(r) dr \quad . \qquad (5)$$

Another atom-in-molecule approach is the Bader charge analysis. It partitions the electron density of a molecule along planes of minimum electron density and assigns maxima at or close to the nuclei to the corresponding atom. Integration over these basins leads to a charge associated with each center.[15]

It is therefore not surprising that different charge assignment schemes would give charges not only differing between different schemes but also different from physically motivated (integer) charge states. In some cases, assignment of computed charges can reliably be made to charge states significantly different from nominal charges returned by Mulliken or Bader charges, see e.g. Refs. 24, 25. Should the nominal computed charges of Ti in $TiO_2$, which are very different from +4, be still assigned to the +4 oxidation state? In this work, we answer in the negative and argue that the assumption of the +4 oxidation state should instead be revised.

The structure of this work is as follows: first, spherically averaged densities of a single Ti atom in different oxidation states are compared among each other and with different methods for a comparison of the valence densities obtained from DFT and wave function-based methods and to determine characteristic features of the density distribution, such as evidence of the nuclear cusp. Second, spherically averaged densities for single $TiO_2$ molecules are analyzed and different charge analyses applied, again with different methods, to quantify the charge remainder at the Ti center. Last, density distributions and charge remainders are computed for the periodic titania systems rutile and anatase to give an estimation of the atomic charge of Ti in solid $TiO_2$ compounds. Additionally, cumulative charges obtained as charge density integrated over radial distance from the Ti centers were computed to verify the charge analysis schemes and to relate the spherically averaged density profile and the band structure to the Ti oxidation state in titania. We conclude that the oxidation state is not +4 and likely +3.

## 2. Methods

Calculations on atoms and molecules were performed using the computational chemistry program package GAUSSIAN 09.[26] For coupled cluster singles and doubles (CCSD)[27,28] and complete active space self-consistent field (CAS-SCF) as well as density functional theory calculations with PBE[29,30] and B3LYP[31-33] functionals, Dunning's polarized valence double-zeta basis set (*cc*-pVDZ)[34-36] was used. Molecular geometries were optimized using B3LYP and the resulting geometries were employed for all molecular calculations. CAS-SCF calculations were performed including Ti 4*s* and 3*d* levels, as well as additionally O 2*s* and 2*p* levels for $TiO_2$ molecules.

Valence electron densities for all methods were extracted from the results and integrated over both angular coordinates around the Ti center to obtain radial densities along the Ti-O bonds for molecules. Similar radial density distributions were also produced for atoms and ions.

To exclude the possibility of a significant basis set influence on the charge analysis, the charges were computed with larger augmented and non-augmented *cc*-pVXZ (with X=2,3,4,5) basis sets and an extrapolation was done to obtain the approximate Bader charge at the basis set limit for molecular $TiO_2$ using a function of the form $f(x)=ax^{-b}+c$.[37,38]

Full-potential all-electron computations on periodic rutile and anatase structures were done with the FHI-aims package using atom-centered orbitals, employing the PBE functional.[39-42] An 8x8x12 k-point grid for rutile and a 12x12x4 grid for anatase were utilized, using a unit cell and the standard basis set definition up to first tier for Ti and second tier for O. Gaussian smearing with a factor of 0.1 and atomic ZORA scalar relativistic treatment were applied. Spin polarization was found to be unimportant. The valence densities were obtained as occupancy-weighted sums of valence orbital densities and integrated in the way previously mentioned.

For a qualitative discussion of the valence density distribution, spherically averaged valence densities $\bar{\rho}_v(r)$ were plotted against distance from the titanium atom, obtained by pointwise division of the radially dependent valence densities $\rho(r)$ by the surface areas of spheres with radius r at each radial point. Furthermore, Mulliken and Hirshfeld population analyses as well as a grid-based Bader charge analyses using the BADER v0.95a code[43-46] were applied to the molecular and periodic systems.

## 3. Results and discussion

*3.1 Valence density distributions in $Ti^{n+}$ (n=0,1,2,3,4)*

Spherically averaged valence densities for a single Ti center in different oxidation states are shown in Fig. 1 for different methods, their lowest-energy valence configuration is indicated after the method name. The expected ground states for Ti, $Ti^+$, $Ti^{2+}$ and $Ti^{3+}$ are $^3F$ (valence configuration $3d^24s^2$), $^4F$ ($3d^24s^1$), $^3F$ ($3d^24s^0$) and $^2D$ ($3d^14s^0$).[47,48] As can be seen from Fig. 1, CAS-SCF is the only method which is capable of predicting the correct ground states of both, $Ti^+$ and $Ti^{2+}$, due to the multiconfigurational character of the wave functions. Also in Fig. 1 the cumulative charge around $Ti^{n+}$ within a sphere of a given radius is shown; this charge converges to 4, 3, 2, and 1 $e$ for $Ti^0$, $Ti^+$, $Ti^{2+}$ and $Ti^{3+}$, respectively.

Two qualitative features of the spherically averaged radial valence density can be identified with all methods: first, a local maximum of electron density at a radial distance of approximately 0.2-0.4 Å from the nucleus which corresponds to *d*-type contributions since it is present even in the absence of *s*-type valence density; second, a local maximum of the averaged valence density close to the Ti nuclear position, whose occurrence is connected to remaining *s*-contributions, since the radial node of basis functions with $l > 0$ leads to a steep decrease and vanishing of *d*-type density close to the nucleus. The cumulative charges show that a significant fraction of the overall charge is contained in a sphere including the *s*- and *d*-type peaks and it integrates to unity between 0 and 0.6 or 0.9 Å for $Ti^0$ or $Ti^{2+}$, respectively. The lack of valence electrons in $Ti^{4+}$ results in a zero average valence density over the whole space. In the case of an eventually occurring, fully oxidized Ti center in $TiO_2$ compounds, no peak at r→0 or a peak between 0.2-0.4 Å is expected.

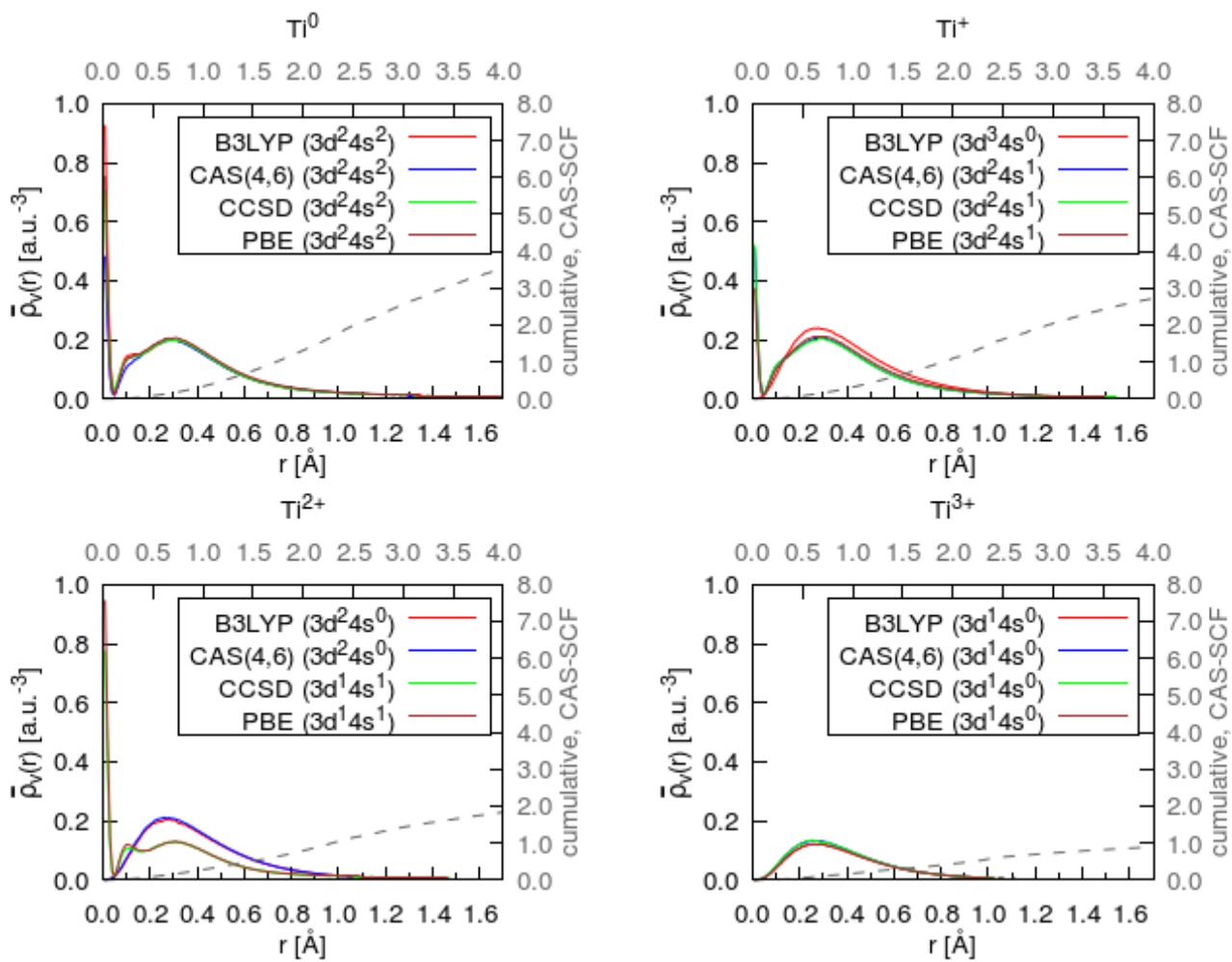

*Figure 1:* Spherically averaged valence electron density of neutral and cationic Ti species obtained with B3LYP, CAS(4,6), CCSD and PBE with the *cc*-pVDZ basis set. The spatial coordinate r corresponds to the radial distance from the Ti center. The gray, dashed lines indicate the cumulative numbers of electrons within the spheres of corresponding radius (which converge at r→∞ to 4, 3, 2, and 1 *e* for $Ti^0$, $Ti^+$, $Ti^{2+}$ and $Ti^{3+}$ respectively).

*3.2 Valence densities of $TiO_2$ molecules*

The lowest-energy geometry of a $TiO_2$ molecule is a bent structure with a Ti-O distance of 1.635 Å and a bond angle of 111.75°. Since O-Ti-O angles in rutile are either 90° or 180° and in anatase around 77°, 102° and 180°, we decided to include a linear molecular species as another model system as well. Although not a local minimum of the potential energy surface (PES), a linear structure with a Ti-O distance of 1.699 Å, lying about 2 eV above the bent species, was identified as a saddle point on the molecular PES. The obtained parameters are in good agreement with previously reported molecular geometries of $TiO_2$.[49] Fig. 2 shows the spherically averaged radial valence density for both geometries and different methods.

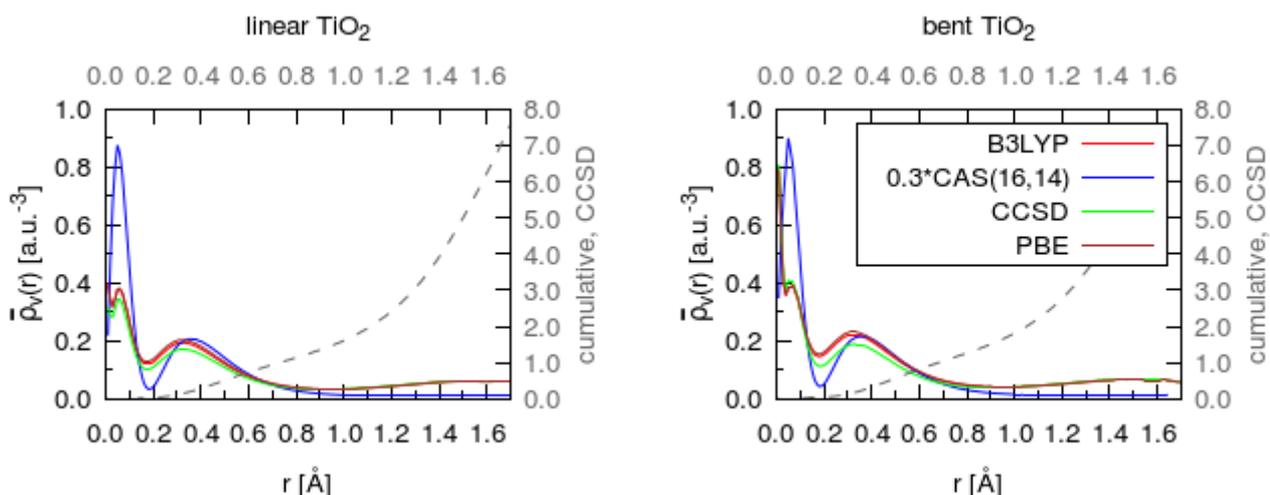

*Figure 2:* Spherically averaged valence electron density of a linear and bent $TiO_2$ molecule obtained with B3LYP, CAS(14,16), CCSD and PBE with the *cc*-pVDZ basis set. The spatial coordinate r corresponds to the radial distance from the Ti center. The gray, dashed lines indicate the cumulative numbers of electrons within the spheres of corresponding radius. The CAS(16,14) curve is scaled by a factor of 0.3 for a better comparison.

As can be seen in Fig. 2, a local maximum of the density in the vicinity of the Ti nucleus as well as a local maximum around 0.3 Å from Ti occurs in both cases and for all used methods. This is similar to the case of Ti ions shown in Fig.1. Also similarly to the case of Ti ions, the cumulative charge reaches unity around 0.7 Å, including most of the identified Ti *s*- and *d*-type density contributions, suggesting one remaining Ti electron at the metal center within a sphere with a radius of approximately 40-43% of the Ti-O bond length.

The resulting curve in Fig. 2 for CAS-SCF is scaled by a factor of 0.3 for a better comparison. The CAS(16,14) values show noticeable deviations from the other methods, predicting an even larger density remainder at the Ti atom, indicating again a strong influence of static correlation which is expected to be lower in the octahedrally coordinated Ti in periodic systems. Despite the obvious quantitative deviations, the CAS-SCF curve shows the above mentioned characteristic points, the pronounced cusp-like maximum at r=0 is not depicted due to the grid resolution.

Remaining electron density on Ti is also predicted by the three types of charge analyses chosen, as shown in Table 1.

Table 1: Mulliken, Bader and Hirshfeld charges on titanium for the linear and bent configurations of the TiO$_2$ molecule obtained with B3LYP, CAS(16,14), CCSD and PBE with the *cc*-pVDZ basis set.

| System/Method | | Mulliken | Bader | Hirshfeld |
|---|---|---|---|---|
| **linear** | *B3LYP* | +0.796 | +2.445 | +0.989 |
| | *CAS(16,14)* | +0.859 | +2.023 | +1.117 |
| | *CCSD* | +0.790 | +2.610 | +1.023 |
| | *PBE* | +0.732 | +2.342 | +0.927 |
| **bent** | *B3LYP* | +0.661 | +2.149 | +0.864 |
| | *CAS(16,14)* | +0.742 | +1.878 | +0.893 |
| | *CCSD* | +0.670 | +2.252 | +0.899 |
| | *PBE* | +0.594 | +2.052 | +0.798 |

The charge measures vary significantly among different charge analysis methods while they agree well between different electronic structure methods used for the same charge analysis method. For oxygen the respective values obtained were approximately -0.3 with Mulliken, -1.1 with Bader and -0.5 with Hirshfeld charge analysis.

To exclude the possibility of the shown valence density increase at the Ti center to be predominantly caused by diffuse O-centered functions, a spherically averaged valence density plot of an oxygen dimer in the same relative alignment as in the linear species is shown (computed with CCSD) in Fig. 3 for different reduced states (O$_2$, O$_2^-$, O$_2^{2-}$, O$_2^{3-}$, O$_2^{4-}$), with r=0 being at the center of the elongated O$_2$ molecule or molecule-ions (where the Ti atom would have been in a linear TiO$_2$ molecule).

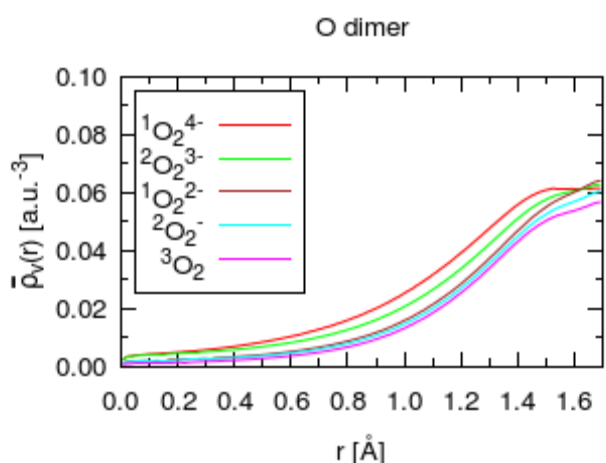

*Figure 3:* Spherically averaged valence electron density of O$_2^{n-}$ (n=0,1,2,3,4) with a bond length of R=3.398 Å and the point r=0 at the center of the bond. The valence densities were calculated with CCSD/*cc*-pVDZ.

In Fig. 3 can be seen that the spherically averaged valence density of the O dimer is decreasing towards the center of the bond and not showing any peaks around r=0 for each oxidation state. This leads to the conclusion that the increase in valence density at the Ti center in $TiO_2$ is indeed caused by remaining charge density on titanium.

The Mulliken charges, although expected to be the least quantitatively reliable method, indicate significant contributions of Ti-centered basis functions to the electron density as already expected from the averaged radial valence density plots. The Hirshfeld analysis, based on the ratio of the individual spherically averaged atomic densities (as the ones shown in Fig. 1 and Fig. 3) and their sum multiplied by the molecular density, as shown in eq. (5), leads to slightly larger positive charges on Ti, but still significantly lower than the expected +4 oxidation state.

The Bader charge analysis predicts the lowest value for the remaining charge on the Ti center of all methods. The minimum distance of the dividing surface to the titanium atom is 0.69 Å for CCSD and 0.72 Å for B3LYP and PBE in the case of a linear molecule and 0.60 Å for CCSD as well as 0.59 Å for B3LYP and PBE in the bent molecule. Comparing these distances to the average density plots in Fig. 2 makes clear that the Bader analysis includes the parts of the density previously identified as characteristic of the Ti center. The Bader maxima assigned to the Ti atoms are also in agreement with the radial maxima belonging to Ti shown in Fig. 2, suggesting a correct partitioning of the electron density within the grid-based Bader charge analysis formalism. Since CCSD, B3LYP and PBE results do not differ significantly regarding density plots and charge analyses, PBE is assumed to be accurate for the purpose of charge analysis and is used in the calculations on periodic systems discussed in the next section.

The Bader charge remainder at Ti in the linear $TiO_2$ molecule obtained with PBE and larger (aug-)*cc*-pVXZ (X=2,3,4,5) basis sets are plotted in Fig. 4 and extrapolated to the basis set limit. For both basis set series the remaining charges at Ti converge to the same limit of +2.448 which corresponds to a deviation from the PBE/*cc*-pVDZ value of 4.3%. Although there is a small basis set influence, it can be stated that charges retrieved from the smaller basis set are in good agreement with the expected basis set limit.

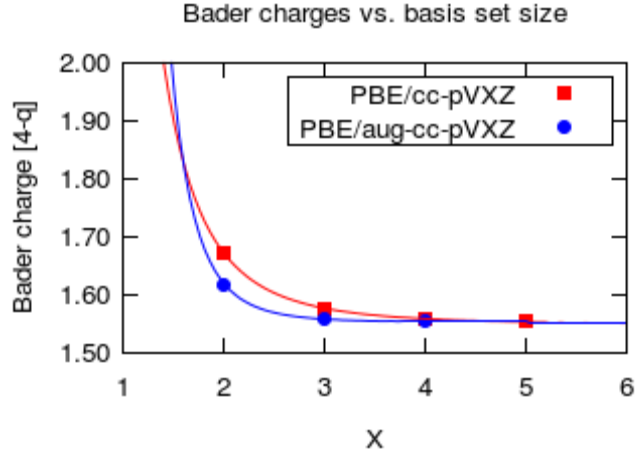

*Figure 4:* Remaining charge (4-q, where q is the Bader charge of Ti) on Ti in a linear $TiO_2$ molecule obtained by Bader analysis with PBE and augmented and non-augmented *cc*-pVXZ basis sets, where X=2,3,4,5 (points) and corresponding power function fit (lines).

*3.3 Rutile and anatase*

The computed structures of anatase and rutile $TiO_2$ are in good agreement with available experimental and theoretical values: a=3.81 Å (experimental[50]: 3.79 Å, theoretical (GGA)[51]: 3.81 Å) and c=9.72 Å (9.51 Å, 9.69 Å) for anatase, a=4.65 Å (4.59 Å, 4.65 Å) and c=2.97 Å (2.96 Å, 2.97 Å) for rutile. The structures were visualized using the software VESTA[52] and are shown in Fig. 5.

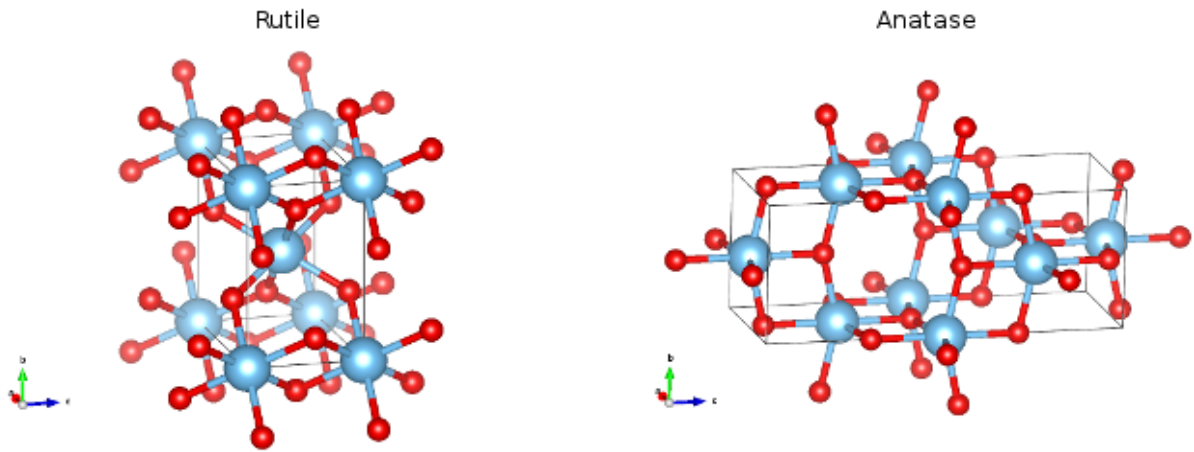

*Figure 5:* Optimized crystal structures of rutile (left) and anatase (right). Depicted are the unit cells used throughout this work, the blue spheres indicate the Ti, red spheres the O positions.

The Ti-O distances in the periodic structures are 1.96 Å in rutile as well as 1.95 Å and 2.01 Å in anatase. The spherically averaged valence densities around Ti for both systems are shown in

Fig. 6.

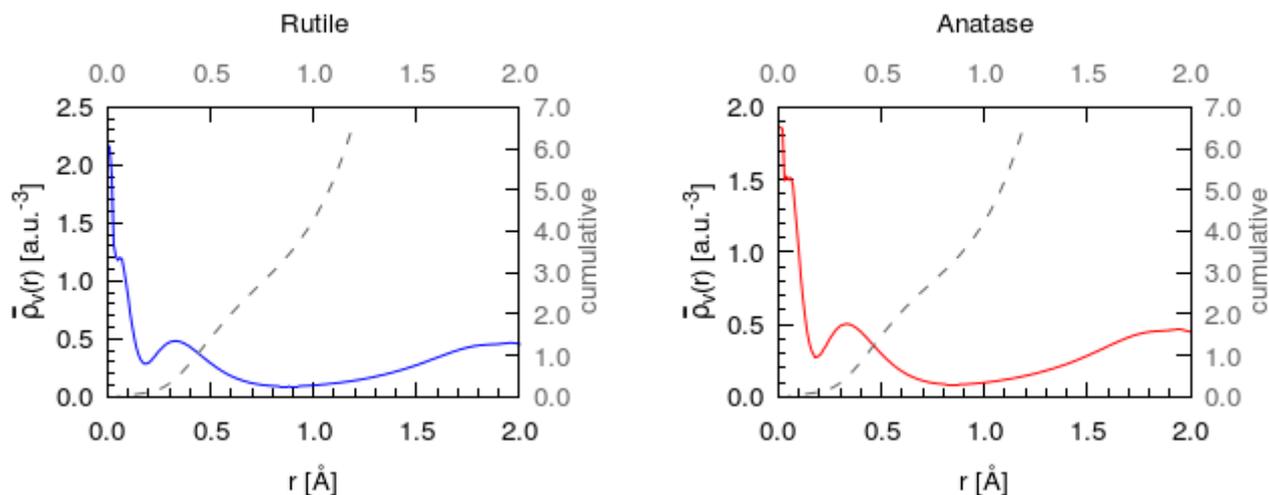

*Figure 6:* Spherically averaged valence electron density around the Ti atom of a periodic rutile and anatase structure calculated with PBE. The spatial coordinate r corresponds to the radial distance from one Ti center. The O atoms are located at r=1.96 Å in rutile and r=1.95 Å and 2.01 Å in anatase. The gray, dashed lines indicate the cumulative numbers of electrons within the spheres of corresponding radius.

Once again, the plot exhibits a clear peak of $\bar{\rho}_v(r)$ at the Ti position and a local maximum in its proximity (specifically, in the area of r=0.2...0.5 Å identified above as coming from $3d$ Ti valence electrons), suggesting significant valence electron density remaining at the metal atom. The cumulative charge integrates to unity around 0.44 Å, including *s*- and *d*-type contributions, as in the molecule, but closer to the nucleus due to the increased charge density around the metal center caused by a larger number of coordinating oxygen ligands. Qualitatively, the picture does not change and the data suggest one remaining electron within a small sphere confined to the metal center and a radius of now approximately 22% of the Ti-O distance.

Similarly to the molecular case, Mulliken, Hirshfeld and Bader population analyses were used on the periodic systems and the results are summarized in Table 2.

*Table 2:* Mulliken, Bader and Hirshfeld charge analyses for a Ti atom in the periodic $TiO_2$ structures rutile and anatase calculated with PBE.

| System | Mulliken | Bader | Hirshfeld |
|---|---|---|---|
| *Rutile* | +0.646 | +2.521 | +0.574 |
| *Anatase* | +0.649 | +2.495 | +0.554 |

The results differ only slightly from the molecular case with the values of Mulliken charges of Ti atoms in the crystals being between those of linear and bent configurations of the $TiO_2$ molecules, Hirshfeld charges being lower, Bader charges larger than for the molecules. The corresponding charges for the oxygen atoms are approximately -0.3 with Mulliken and Hirshfeld as well as -1.3 with Bader charge analysis. However, once again the data suggest a substantial remaining charge on the Ti atom which does not align with the conventional perception of the +4 oxidation state of Ti in titanium dioxide compounds.

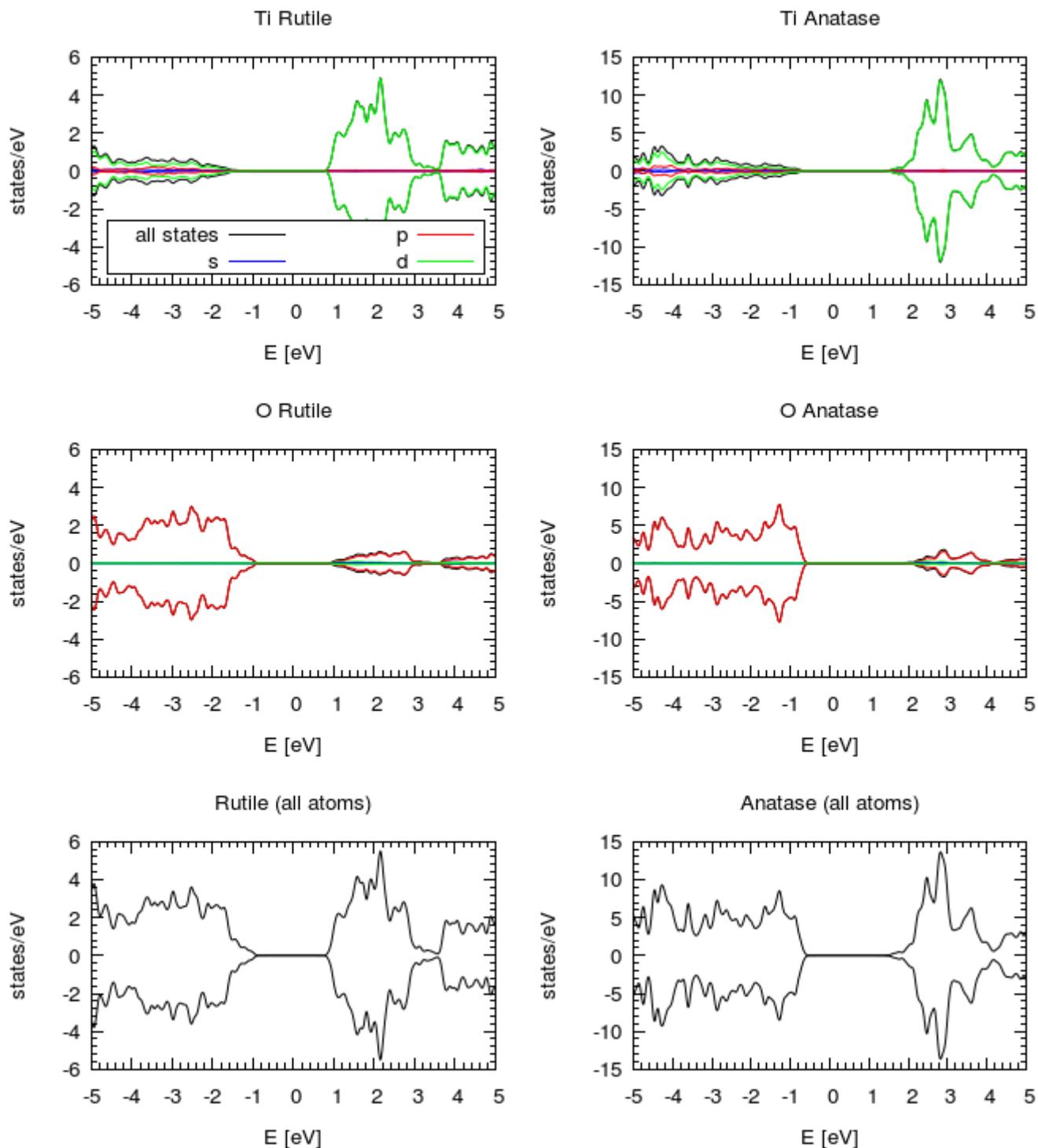

*Figure 7:* *l*-projected partial density density of states for Ti and O in rutile and anatase as well as DOS for all atoms and states.

This is in agreement with the observation that the TiO$_2$ valence band has significant Ti *d*-contributions as can be seen in the *l*-projected partial density of states shown in Fig. 7 for Ti and O in rutile and anatase. Fig. 7 is in agreement with pDOS plots of titania published in extensive theoretical literature.[6,8,9] Also, similar results to those presented here regarding DOS and charge analyses can be obtained by means of plane wave basis sets and using pseudopotentials.[53,16]

Valence density distribution, cumulative charge and the employed charge analysis schemes suggest, in combination with the *l*-projected pDOS, an oxidation state of +3 for Ti. The respective oxidation state of oxygen would therefore correspond to -1.5.

## 4. Conclusion

The commonly used charge analysis schemes suggest that the titanium oxidation state in molecular and solid TiO$_2$ is between +0.6 to +2.6. The Bader charge analysis can be considered as the most reliable one in this sequence, since it directly evaluates the electron density remainder around the nucleus and is therefore also directly related to the actual charge density; furthermore the positions of zero-flux surfaces are in good agreement with the distance at which the cumulative charges integrate to unity. It is common to assign integer oxidation states as corresponding to fractional computed charges. In many instances, there is a clear justification of such an assignment, based for example on the analysis of the band structure.[25,54] In the case of TiO$_2$ however, there is no solid justification for assigning e.g. the Bader charge of about +2.5 *e* to an oxidation state Ti$^{4+}$. Yet such an oxidation state has been universally assumed in the literature. Here, we performed a comprehensive analysis of the electronic structure of Ti ions, TiO$_2$ molecules and TiO$_2$ solids which indicates that a significant valence charge remains on Ti corresponding to an oxidation state of +3 in TiO$_2$ molecules as well as in anatase and rutile crystals. This follows from the persistence of the nuclear cusp-like maximum due to *s* valence states of Ti on the Ti center in TiO$_2$. We also showed that the peak in electron density of a Ti atom at ~0.2-0.5 Å away from the core, which is due to the valence *d* states of Ti, persists in TiO$_2$ as well. The amount of charge due to the cusp-like and *d*-type density maximum amounts to about one electron in TiO$_2$ molecules and crystals which is consistent with the isolated Ti ions.

This suggests a revision of the assumed Ti oxidation state in TiO$_2$, at least in the framework of a computational treatment. This is of particular interest in the case of redox processes in TiO$_2$ systems; for example the assignment of oxidation states obtained in doped TiO$_2$ is done off the assumed +4 oxidation state in pure TiO$_2$.[55-57] Our results suggest that, in principle, further oxidation at the Ti centers is possible. Conversely follows that oxygen could be further reduced, oxygen redox activity is a known effect occurring e.g. upon Li intercalation in several cathode materials.[58]

## 5. Acknowledgements

This work was supported by the Ministry of Education of Singapore (grant no. MOE2015-T2-1-011).

## 6. References

(1) Nakata, K.; Fujishima, A. TiO$_2$ Photocatalysis: Design and Applications. *J. Photochem. Photobiol. C Photochem. Rev.* **2012**, *13* (3), 169–189.
(2) J. Frank, A.; Kopidakis, N.; Lagemaat, J. van de. Electrons in Nanostructured TiO$_2$ Solar Cells: Transport, Recombination and Photovoltaic Properties. *Coord. Chem. Rev.* **2004**, *248* (13–14), 1165–1179.
(3) Schneider, J.; Matsuoka, M.; Takeuchi, M.; Zhang, J.; Horiuchi, Y.; Anpo, M.; Bahnemann, D. W. Understanding TiO$_2$ Photocatalysis: Mechanisms and Materials. *Chem. Rev.* **2014**, *114* (19), 9919–9986.
(4) Liu, J.; Wang, J.; Ku, Z.; Wang, H.; Chen, S.; Zhang, L.; Lin, J.; Shen, Z. X. Aqueous Rechargeable Alkaline Co$_x$Ni$_{2-x}$S$_2$/TiO$_2$ Battery. *ACS Nano* **2016**, *10* (1), 1007–1016.
(5) Oh, S.-M.; Hwang, J.-Y.; Yoon, C. S.; Lu, J.; Amine, K.; Belharouak, I.; Sun, Y.-K. High Electrochemical Performances of Microsphere C-TiO$_2$ Anode for Sodium-Ion Battery. *ACS Appl. Mater. Interfaces* **2014**, *6* (14), 11295–11301.
(6) Mo, S.-D.; Ching, W. Y. Electronic and Optical Properties of Three Phases of Titanium Dioxide: Rutile, Anatase, and Brookite. *Phys. Rev. B* **1995**, *51* (19), 13023–13032.
(7) Glassford, K. M.; Chelikowsky, J. R. Structural and Electronic Properties of Titanium Dioxide. *Phys. Rev. B* **1992**, *46* (3), 1284–1298.
(8) Scanlon, D. O.; Dunnill, C. W.; Buckeridge, J.; Shevlin, S. A.; Logsdail, A. J.; Woodley, S. M.; Catlow, C. R. A.; Powell, M. J.; Palgrave, R. G.; Parkin, I. P.; et al. Band Alignment of Rutile and Anatase TiO$_2$. *Nat Mater* **2013**, *12* (9), 798–801.
(9) Rubio-Ponce, A.; Conde-Gallardo, A.; Olguín, D. First-Principles Study of Anatase and Rutile TiO$_2$ Doped with Eu Ions: A Comparison of GGA and LDA+U Calculations. *Phys. Rev. B* **2008**, *78* (3), 35107.
(10) Rogers, D. B.; Shannon, R. D.; Sleight, A. W.; Gillson, J. L. Crystal Chemistry of Metal Dioxides with Rutile-Related Structures. *Inorg. Chem.* **1969**, *8* (4), 841–849.
(11) Tossell, J. A.; Vaughan, D. J. Electronic Structure of Rutile, Wustite, and Hematite from Molecular Orbital Calculations. *Am Miner.* **1974**, *59* (3–4), 319–334.
(12) Scrocco, M. X-Ray Photoelectron Spectra of Ti$^{4+}$ in TiO$_2$. Evidence of Band Structure. *Chem. Phys. Lett.* **1979**, *61* (3), 453–456.
(13) Local Bonding Analysis of the Valence and Conduction Band Features of TiO$_2$. *J. Appl. Phys.* **2007**, *102* (3), 33707.
(14) Mulliken, R. S. Electronic Population Analysis on LCAO–MO Molecular Wave Functions. I. *J. Chem. Phys.* **1955**, *23* (10), 1833–1840.
(15) Bader, R. F. W. Atoms in Molecules. In *Encyclopedia of Computational Chemistry*; John Wiley & Sons, Ltd, 2002.
(16) Arrouvel, C.; Parker, S. C.; Islam, M. S. Lithium Insertion and Transport in the TiO$_2$−B

Anode Material: A Computational Study. *Chem. Mater.* **2009**, *21* (20), 4778–4783.
(17)  Daude, N.; Gout, C.; Jouanin, C. Electronic Band Structure of Titanium Dioxide. *Phys. Rev. B* **1977**, *15* (6), 3229–3235.
(18)  Chauque, S.; Robledo, C. B.; Leiva, E. P. M.; Oliva, F. Y.; Cámara, O. R. Comparative Study of Different Alkali (Na, Li) Titanate Substrates as Active Materials for Anodes of Lithium - Ion Batteries. *ECS Trans.* **2014**, *63* (1), 113–128.
(19)  Wu, L.; Bresser, D.; Buchholz, D.; Giffin, G. A.; Castro, C. R.; Ochel, A.; Passerini, S. Unfolding the Mechanism of Sodium Insertion in Anatase $TiO_2$ Nanoparticles. *Adv. Energy Mater.* **2015**, *5* (2), 1401142.
(20)  Kato, T. On the Eigenfunctions of Many-Particle Systems in Quantum Mechanics. *Commun. Pure Appl. Math.* **1957**, *10* (2), 151–177.
(21)  Hirshfeld, F. L. Bonded-Atom Fragments for Describing Molecular Charge Densities. *Theor. Chim. Acta* **1977**, *44* (2), 129–138.
(22)  Ritchie, J. P.; Bachrach, S. M. Some Methods and Applications of Electron Density Distribution Analysis. *J. Comput. Chem.* **1987**, *8* (4), 499–509.
(23)  Ritchie, J. P. Electron Density Distribution Analysis for Nitromethane, Nitromethide, and Nitramide. *J. Am. Chem. Soc.* **1985**, *107* (7), 1829–1837.
(24)  Legrain, F.; Manzhos, S. Aluminum Doping Improves the Energetics of Lithium, Sodium, and Magnesium Storage in Silicon: A First-Principles Study. *J. Power Sources* **2015**, *274*, 65–70.
(25)  Liu, W.; Sk, M. A.; Manzhos, S.; Agarwal, J.; Schaefer III, H. F.; Carrington, T. Grown-in Beryllium Diffusion in Indium Gallium Arsenide: An Ab Initio, Continuum Theory and Kinetic Monte Carlo Study. *Acta Mater. accepted*.
(26)  Gaussian 09, Revision A.02, M. J. Frisch, G. W. Trucks, H. B. Schlegel, G. E. Scuseria, M. A. Robb, J. R. Cheeseman, G. Scalmani, V. Barone, G. A. Petersson, H. Nakatsuji, X. Li, M. Caricato, A. Marenich, J. Bloino, B. G. Janesko, R. Gomperts, B. Mennucci, H. P. Hratchian, J. V. Ortiz, A. F. Izmaylov, J. L. Sonnenberg, D. Williams-Young, F. Ding, F. Lipparini, F. Egidi, J. Goings, B. Peng, A. Petrone, T. Henderson, D. Ranasinghe, V. G. Zakrzewski, J. Gao, N. Rega, G. Zheng, W. Liang, M. Hada, M. Ehara, K. Toyota, R. Fukuda, J. Hasegawa, M. Ishida, T. Nakajima, Y. Honda, O. Kitao, H. Nakai, T. Vreven, K. Throssell, J. A. Montgomery, Jr., J. E. Peralta, F. Ogliaro, M. Bearpark, J. J. Heyd, E. Brothers, K. N. Kudin, V. N. Staroverov, T. Keith, R. Kobayashi, J. Normand, K. Raghavachari, A. Rendell, J. C. Burant, S. S. Iyengar, J. Tomasi, M. Cossi, J. M. Millam, M. Klene, C. Adamo, R. Cammi, J. W. Ochterski, R. L. Martin, K. Morokuma, O. Farkas, J. B. Foresman, and D. J. Fox, Gaussian, Inc., Wallingford CT, 2016.
(27)  Scuseria, G. E.; Janssen, C. L.; Schaefer III, H. F. An Efficient Reformulation of the Closed-shell Coupled Cluster Single and Double Excitation (CCSD) Equations. *J. Chem. Phys.* **1988**, *89* (12), 7382–7387.
(28)  Čížek, J. On the Use of the Cluster Expansion and the Technique of Diagrams in Calculations of Correlation Effects in Atoms and Molecules. In *Advances in Chemical Physics*; John Wiley & Sons, Inc., 1969; pp 35–89.
(29)  Perdew, J. P.; Burke, K.; Ernzerhof, M. Generalized Gradient Approximation Made Simple [Phys. Rev. Lett. 77, 3865 (1996)]. *Phys. Rev. Lett.* **1997**, *78* (7), 1396–1396.
(30)  Perdew, J. P.; Burke, K.; Ernzerhof, M. Generalized Gradient Approximation Made Simple. *Phys. Rev. Lett.* **1996**, *77* (18), 3865–3868.
(31)  Lee, C.; Yang, W.; Parr, R. G. Development of the Colle-Salvetti Correlation-Energy Formula into a Functional of the Electron Density. *Phys. Rev. B* **1988**, *37* (2), 785–789.
(32)  Density-functional Thermochemistry. III. The Role of Exact Exchange. *J. Chem. Phys.* **1993**, *98* (7), 5648–5652.
(33)  Devlin, F. J.; Finley, J. W.; Stephens, P. J.; Frisch, M. J. Ab Initio Calculation of Vibrational Absorption and Circular Dichroism Spectra Using Density Functional Force Fields: A Comparison of Local, Nonlocal, and Hybrid Density Functionals. *J. Phys. Chem.* **1995**, *99* (46), 16883–16902.

(34) Balabanov, N. B.; Peterson, K. A. Systematically Convergent Basis Sets for Transition Metals. I. All-Electron Correlation Consistent Basis Sets for the 3d Elements Sc–Zn. *J. Chem. Phys.* **2005**, *123* (6), 64107.
(35) Balabanov, N. B.; Peterson, K. A. Basis Set Limit Electronic Excitation Energies, Ionization Potentials, and Electron Affinities for the 3d Transition Metal Atoms: Coupled Cluster and Multireference Methods. *J. Chem. Phys.* **2006**, *125* (7), 74110.
(36) Dunning Jr., T. H. Gaussian Basis Sets for Use in Correlated Molecular Calculations. I. The Atoms Boron through Neon and Hydrogen. *J. Chem. Phys.* **1989**, *90* (2), 1007–1023.
(37) Basis-Set Convergence of Correlated Calculations on Water. *J. Chem. Phys.* **1997**, *106* (23), 9639–9646.
(38) Klopper, W.; Helgaker, T. Extrapolation to the Limit of a Complete Basis Set for Electronic Structure Calculations on the $N_2$ Molecule. *Theor. Chem. Acc.* **1998**, *99* (4), 265–271.
(39) Blum, V.; Gehrke, R.; Hanke, F.; Havu, P.; Havu, V.; Ren, X.; Reuter, K.; Scheffler, M. Ab Initio Molecular Simulations with Numeric Atom-Centered Orbitals. *Comput. Phys. Commun.* **2009**, *180* (11), 2175–2196.
(40) Havu, V.; Blum, V.; Havu, P.; Scheffler, M. Efficient Integration for All-Electron Electronic Structure Calculation Using Numeric Basis Functions. *J. Comput. Phys.* **2009**, *228* (22), 8367–8379.
(41) Ren, X.; Rinke, P.; Blum, V.; Wieferink, J.; Tkatchenko, A.; Sanfilippo, A.; Reuter, K.; Scheffler, M. Resolution-of-Identity Approach to Hartree–Fock, Hybrid Density Functionals, RPA, MP2 and GW with Numeric Atom-Centered Orbital Basis Functions. *New J. Phys.* **2012**, *14* (5), 53020.
(42) Marek, A.; Blum, V.; Johanni, R.; Havu, V.; Lang, B.; Auckenthaler, T.; Heinecke, A.; Bungartz, H.-J.; Lederer, H. The ELPA Library: Scalable Parallel Eigenvalue Solutions for Electronic Structure Theory and Computational Science. *J. Phys. Condens. Matter* **2014**, *26* (21), 213201.
(43) Tang, W.; Sanville, E.; Henkelman, G. A Grid-Based Bader Analysis Algorithm without Lattice Bias. *J. Phys. Condens. Matter* **2009**, *21* (8), 84204.
(44) Henkelman, G.; Arnaldsson, A.; Jónsson, H. A Fast and Robust Algorithm for Bader Decomposition of Charge Density. *Comput. Mater. Sci.* **2006**, *36* (3), 354–360.
(45) Sanville, E.; Kenny, S. D.; Smith, R.; Henkelman, G. Improved Grid-Based Algorithm for Bader Charge Allocation. *J. Comput. Chem.* **2007**, *28* (5), 899–908.
(46) Yu, M; Trinkle, D. R. Accurate and Efficient Algorithm for Bader Charge Integration. *J. Chem. Phys.* **2011**, *134* (6), 64111.
(47) Saloman, E. B. Energy Levels and Observed Spectral Lines of Neutral and Singly Ionized Titanium, Ti I and Ti II. *J. Phys. Chem. Ref. Data* **2012**, *41* (1), 013101-01-013101–116.
(48) Sugar, J.; Corliss, C. *Atomic Energy Levels of the Iron-Period Elements: Potassium through Nickel*; United States, 1985; p 680.
(49) Lin, C.-K.; Li, J.; Tu, Z.; Li, X.; Hayashi, M.; Lin, S. H. A Theoretical Search for Stable Bent and Linear Structures of Low-Lying Electronic States of the Titanium Dioxide ($TiO_2$) Molecule. *RSC Adv.* **2011**, *1* (7), 1228–1236.
(50) Cromer, D. T.; Herrington, K. The Structures of Anatase and Rutile. *J. Am. Chem. Soc.* **1955**, *77* (18), 4708–4709.
(51) DFT+U Calculations of Crystal Lattice, Electronic Structure, and Phase Stability under Pressure of $TiO_2$ Polymorphs. *J. Chem. Phys.* **2011**, *135* (5), 54503.
(52) Momma, K.; Izumi, F. VESTA 3 for Three-Dimensional Visualization of Crystal, Volumetric and Morphology Data. *J. Appl. Crystallogr.* **2011**, *44* (6), 1272–1276.
(53) Zhang, J.; Zhou, P.; Liu, J.; Yu, J. New Understanding of the Difference of Photocatalytic Activity among Anatase, Rutile and Brookite $TiO_2$. Phys. Chem. Chem. Phys. 2014, 16 (38), 20382–20386.

(54)     Legrain, F.; Manzhos, S. A First-Principles Comparative Study of Lithium, Sodium, and Magnesium Storage in Pure and Gallium-Doped Germanium: Competition between Interstitial and Substitutional Sites. J Chem Phys.

(55)     Di Valentin, C.; Finazzi, E.; Pacchioni, G.; Selloni, A.; Livraghi, S.; Paganini, M. C.; Giamello, E. N-Doped $TiO_2$: Theory and Experiment. Doping Funct. Photoactive Semicond. Met. Oxides 2007, 339 (1–3), 44–56.

(56)     Colón, G.; Maicu, M.; Hidalgo, M. C.; Navío, J. A. Cu-Doped $TiO_2$ Systems with Improved Photocatalytic Activity. Appl. Catal. B Environ. 2006, 67 (1–2), 41–51.

(57)     Morita, K.; Shibuya, T.; Yasuoka, K. Stability of Excess Electrons Introduced by Ti Interstitial in Rutile $TiO_2$(110) Surface. J. Phys. Chem. C 2017, 121 (3), 1602–1607.

(58)     Seo, D.-H.; Lee, J.; Urban, A.; Malik, R.; Kang, S.; Ceder, G. The Structural and Chemical Origin of the Oxygen Redox Activity in Layered and Cation-Disordered Li-Excess Cathode Materials. Nat Chem 2016, 8 (7), 692–697.